\documentstyle[12pt]{article}

\begin{document}

\noindent
\hfill{\small forarxiv/1p1dbox.tex}

\vspace{0.5cm}
\centerline{\Large \bf Untouched Aspects of the Wave Mechanics }

\centerline{\Large \bf of a Particle in One Dimensional Box}

\bigskip
\centerline{\bf Yatendra S. Jain}

\smallskip

\centerline{Department of Physics}

\centerline{ North-Eastern Hill University Shillong - 793 022, India}

\bigskip

\bigskip
\begin{abstract}
Wave mechanics of a particle in 1-D box (size $= d$) is
critically analyzed to reveal its untouched aspects.
When the particle rests in its ground state, its
zero-point force ($F_o$) produces non-zero strain by
modifying the box size from $d$ to $d' = d + \Delta d$
in all practical situations where the force ($F_a$)
restoring $d$ is not infinitely strong.  Assuming that
$F_a$ originates from a potential $\propto x^2$ ($x$
being a small change in $d$), we find that: (i) the
particle and strained box assume a mutually bound state
(under the equilibrium between $F_o$ and $F_a$) with
binding energy $\Delta{E} = -\varepsilon_o'\Delta{d}/d'$
(with $\varepsilon_o' = h^2/8md'^2$ being the ground
state energy of the particle in the strained box),
(ii) the box size oscillates around $d'$ when the said
equilibrium is disturbed, (iii) an exchange of energy
between the particle and the strained box occurs during
such oscillations, and (iv) the particle, having
collisional motion in its excited states, assumes
collisionless motion in its ground state.  These
aspects have desired experimental support and proven
relevance for understanding the physics of widely
different systems such as quantum dots, quantum wires,
trapped single particle/ion, clusters of particles,
superconductors, superfluids, {\it etc.}  It is
emphasized that the physics of such a system in its
low energy states can be truly revealed if the
theory incorporates $F_o$ and related aspects.

\end{abstract}



{\small Email ysjain@email.com}

\bigskip

\centerline{\bf 1. Introduction}

\smallskip
Wave mechanics emerged as a general theory of all
natural phenomena about eight decades back.  While it
is believed to have been successful in explaining the
observed behavior of widely different systems under
widely different physical conditions, its several basic
aspects and their interpretations still form a subject
of debate [1-4].  This reveals that answer to several
philosophical questions related to the basic principles
of wave mechanics are unclear and the intricacies of
wave particle duality need detailed discussion for their
better understanding.  In this context one may consider
even the wave mechanics of a particle of mass $m$ in a
1-D box of size $d$ ({\it the simplest possible quantum
system used to introduce the subject at its elementary
level} [5]) which reveals the following.

\bigskip
\noindent
{\it 1.1} Energy eigenfunction $\Psi_{\rm n}$ of n-th
quantum state
$$\Psi_{\rm n} = \sqrt{\frac{2}{d}}\sin{(q_{\rm n}x})
 \eqno(1)$$

\noindent
with $q_{\rm n} = {\rm n}\pi/d$, n = 1,2,3, ..., energy
eigenvalue,
$$E_{\rm n} =
<\Psi_{\rm n}|-\frac{\hbar^2}{2m}\frac{\partial^2}
{{\partial}x^2}|\Psi_{\rm n}>
= \frac{{\rm n}^2h^2}{8md^2}, \eqno(2)$$

\noindent
with $h=$ Plank constant, and position expectation of
the particle
$$<x> = <\Psi_{\rm n}|x|\Psi_{\rm n}> = \frac{d}{2}
\eqno(3)$$

\bigskip
\noindent
{\it 1.2} $\Psi_{\rm n}$, representing a kind of {\it
stationary matter wave}, modulates the configuration
($x$) and the phase ($\phi_{\rm n} = q_{\rm n}x$)
positions of the particle. It has n anti-nodal regions
of size $\lambda_{\rm n}/2 = d/{\rm n}$ and n-1 nodes
({\it excluding two nodes at the two walls of the box})
and spreads over the entire size ($d$) of the box.  It
is an eigenfunction of $q^2$ operator ($-\partial^2_x$)
not of $q$ ($-i\partial_x$) which means that an
experiment on the particle can determine only the
magnitude (not the direction) of $q$.  The fact, that
$\Psi_{\rm n}$ is a result of the superposition of two
plane waves, $u' = \exp{(iq_{\rm n}x)}$ and $u'' =
\exp{(-iq_{\rm n}x)}$ of momenta $q_{\rm n}$ and
$-q_{\rm n}$, implies that there can be no means that
may fix +/$-$ direction for $q_{\rm n}$ in such a state
of the particle.

\bigskip
\noindent
{\it 1.3} The particle in its n-th quantum state
exerts a {\it real} force [6]
$$f_{\rm n} = - \partial_d\varepsilon_{\rm n}
= \frac{{\rm n}^2h^2}{4md^3} \eqno(4)$$

\noindent
on the walls of the box that tends to increase $d$.

\bigskip
While these results seem to provide full understanding
of the wave mechanics of the system, however, motivated
by our recent analysis of the wave mechanics of two
hard core (HC) particles in 1-D box [7] we analyzed the
wave mechanics of the titled system further and
discovered its untouched aspects, particularly, related
to the role of quantum size, zero-point energy,
zero-point force, {\it etc.}  Interestingly, these
aspects are found to have desired experimental
foundations and proven relevance for understanding of
the true behavior of widely different physical systems
such as quantum dots, quantum wires, trapped ions
[8, 9], metallic clusters [10, 11] and even complex
systems such as superconductors and superfluids
[12, 13].  In this context we may, particularly, refer
to our recent papers related to the microscopic theory
of superconductivity [14] and the unification of the
physics of widely different systems interacting bosons
and fermions including low dimensional systems such as
$N$ HC quantum particles in 1-D box [15].  As such
this study would benefit the learners of wave
mechanics and many body physics.  

\bigskip
\centerline{\bf 2.  Untouched Aspects}

\bigskip
\noindent
{\it 1.1 Strain in the box: }

\bigskip
We note that $f_{\rm n}$ (Eqn. 4) can also be obtained
from
$$f_{\rm n} = 2\hbar q_{\rm n}\frac{\hbar
q_{\rm n}/m}{2d} = \frac{{\rm n}^2h^2}{4md^3}\eqno (5)$$

\noindent
with $2\hbar q_{\rm n}$ = change in momentum of the
particle during its collision with a wall of the box and
$(\hbar q_{\rm n}/m)/(2d)$ = frequency of collision.
This indicates that: (i) the particle between the walls,
moving with group velocity $v_g = \partial E_{\rm n}/
\partial p_{\rm n} = (\hbar q_{\rm n}/m)$, periodically
collides with each wall at a frequency $(\hbar
q_{\rm n}/m)/(2d)$, and (ii) $f_{\rm n}$ has an apparent
identity with the force that a gas particle (as a
classical entity) exerts on a wall of its container and
contributes to the gas pressure which tends to inflate
the size of the container.  Evidently, this is an
interesting aspect of the particle dynamics in a 1-D
box.  In variance with what one learns from classical
mechanics, wave mechanics concludes $E_{\rm n}$ to
depend on $d$ and $E_{\rm 1}$ (lowest possible allowed
energy) to have non-zero value which have an
interesting impact on the behavior of our system which
can be demonstrated by a thought experiment where the
system is kept in contact with a thermal bath whose
temperature ($T$) is slowly reduced to zero.  Since
the probability for the particle to occupy n-th quantum
state goes proportionally with
$\exp{[-(E_{\rm n}-E_1)/k_BT]} = \exp{[-({\rm n}^2-1)
\varepsilon_o/k_BT]}$ (with $E_1 = h^2/8md^2 =
\varepsilon_o$), it is evident that such probability
even for the first excited state (n=2) would be an
order of magnitude smaller than that for the ground
state (n=1) when bath $T$ is $\approx T_o$ (the
temperature equivalent of $\varepsilon_o$).  Evidently,
to a good approximation, the particle at all $T \le
T_o$, stays in its ground state.  Here we note that:
(i) the particle in n $> 1$ state can have lower energy,
not only by inflating $d$, but also by following an
alternative path (where increase in $d$ is not
necessary) of jumping to a lower energy state and (ii)
$f_{\rm n>1}$ decreases in its strength when the
particle moves to lower energy state with decreasing
$T$.  Evidently, the inflation in $d$ (if any) produced
by $f_{\rm n}$ would decrease with decreasing $T$ and
the box would exhibit $+ ve$ thermal expansion
coefficient $(1/d)\partial_Td$.  However, the situation
changes when the particle rests in its ground state
(n = 1) because it can have lower energy only if $d$
gets increased.  Naturally $f_1 (= {h^2}/{4md^3}$),
which can be called as {\it zero-point force}
(hereafter denoted by $F_o$), can produce non-zero
strain ($+ \Delta{d}$ change in $d$) with $T$ decreasing
below $T \approx T_o$ when it reaches in equilibrium
with forces ($F_a$) that restore $d$.  It is evident
that $F_o$ and $+ \Delta{d}$ would remain unchanged at
all $T \le T_o$ and the system would have $- ve$ $(1/d)
\partial_Td$ around $T_o$.  The observation of $- ve$
$(1/d)\partial_Td$ of the system at $T$ around $T_o$
should obviously prove the presence of non-zero strain
in the box  and the fall of the particle in its ground
state.

\bigskip
\noindent
{\it 2.2 Bound state of Particle and strained box }

\bigskip
We note that Eqns. 1-3 assume that the force $F_a$,
which controls the box size, is infinitely strong for
which $d$ remains unchanged in spite of the inflating
action of $f_{\rm n}$.  However, since $F_a$ is not
infinitely strong in all practical situations, $d$ is
expected to change (say, by $\Delta d$) in the state
of equilibrium between $F_a$ and $f_{\rm n}$.
Assuming that $F_a$ can be derived from $V(x) =
(k/2)x^2$ potential, the total energy of the system
(particle in its ground state + strained box) can be
expressed as
$${\rm E} = \frac{h^2}{8m(d + x)^2} +
\frac{1}{2}kx^2 \eqno(6)$$

\noindent
with $x$ being an increase in $d$.  Solving Eqn. 6
for equilibrium where $x = \Delta d$, we have
$$\frac{\partial{\rm E}}{\partial x}|_{x = \Delta d}
 = -\frac{h^2}{4md'^3} + k{\Delta d} = 0 \quad\quad
 {\rm with} \quad d' = d + \Delta d \eqno(7)$$

\noindent
which renders
$$k{\Delta d} = \frac{h^2}{4md'^3} =
\frac{2\varepsilon_o'}{d'} \quad\quad {\rm with} \quad
\varepsilon_o' = \frac{h^2}{8md'^2} \eqno(8)$$

\noindent
Evidently, a change in ${\rm E}$ with a change in $x$
from $x = 0$ to $x = \Delta d$ ({\it i.e.},
$\Delta{\rm E} = {\rm E}_{|x = \Delta d} -
{\rm E}_{|x = 0}$) can be obtained from
$$\Delta{\rm E} = \frac{h^2}{8md'^2} + \frac{1}{2}k{
\Delta d}^2 - \frac{h^2}{8md^2} \approx - \frac{h^2}
{8md'^3}\Delta d = - \frac{\varepsilon_o'
\Delta d}{d'} \eqno(9)$$

\noindent
where we use Eqns. 6 and 8.  Since $\Delta{\rm E}$
represents the net change in energy of the particle
and the strained box as a single unit, its $-ve$ value
signifies that the two assume a kind of single bound
state and they remain in this state unless
$\Delta{\rm E}$ energy is supplied from outside.  We
also note that: (i) strain ($\Delta d$) as well as
strain energy ($k{\Delta d}^2/2$) of the box are the
functions of the ground state energy ($\varepsilon_o'
= h^2/8md'^2$) or momentum ($h/2d'$) of the particle.
In principle, one may observe similar binding (${\Delta
{\rm E}}_{\rm n} = -{\rm n}^2 \varepsilon_{\rm n}
{\Delta d}_{\rm n}/d'$) for any $\Psi_{\rm n}$ state,
however, a real bound state can be observed only for the
particle in $\Psi_{\rm 1}$ state, because the particle in
a $\Psi_{\rm n > 1}$ state is free to jump to any lower
state by releasing out the difference in their energies
which is always $+ve$, while the particle in the ground
state ($\Psi_{\rm 1}$) does not have this option.  It is
for reason that we wrote Eqn. 6 for the particle in its
ground state only. 

\bigskip
\noindent
{\it 2.3 Energy exchange between particle and strained
box}

\bigskip
We note that the total energy of the particle in
$\Psi_1$ state in its equilibrium with strained
box is

$$E_1 =  \frac{h^2}{8md'^2} + \frac{1}{2}k\Delta{d}^2
\eqno(10)$$

\noindent
However, the equilibrium is disturbed if $d'$ is changed
to $d'\pm \eta$ (with $|\eta| < \Delta{d}$) and the
energy of the disturbed system would be
$$E_1' =  \frac{h^2}{8m(d'\pm \eta)^2} +
\frac{1}{2}k(\Delta{d \pm \eta})^2  \eqno(11)$$

\noindent
which can be rearranged as 
$$E_1' \approx \varepsilon_o' +  \epsilon_s
\mp (\frac{2\varepsilon_o'}{d'} - k\Delta{d})\eta
+ \frac{1}{2}k'\eta^2         \eqno(12)$$

\noindent
with $\epsilon_s = \frac{1}{2}k\Delta{d}^2$ being the
strain energy of the box and modified force constant 
$$k' =  k + \frac{6\varepsilon_o}{d'^2}  \eqno(13)$$

\noindent
Since the first two terms on the right hand side of
Eqn. 12 are constant and $(2\varepsilon_o'/d' - k
\Delta{d})\eta$ vanishes for the equilibrium condition
(Eqn. 8), it can be argued that, with the particle
occupying its ground state energy, the box size
oscillations are controlled by an increased force
constant $k'$ (Eqn. 13).  In addition
$(2\varepsilon_o'/d' - k\Delta{d})\eta = 0$ implies that
zero point energy of the particle increases (or
{\it decreases}) by $(2\varepsilon_o'\eta/d')$ when
strain energy decreases ( or {\it increases}) by equal
amount, ($k\Delta{d})\eta$) indicating that the particle
and the strained box keeps exchanging energy with each
other during $\eta$ oscillations.   

\bigskip
\noindent
{\it 2.4 Self superposition, effective size and
nature of motion:}

\bigskip
Since $\Psi_{\rm n}$ is the superposition of two plain
waves ({\it viz.,} $\exp{(iqx)}$ and its reflection
$\exp{(-iqx)}$ representing one and the same particle),
the particle in the box assumes a kind of self
superposition state.  Identifying these waves to
represent two separate particles, one may find that
$\Psi_{\rm n}$ can identically define the state of
mutual superposition of two particles moving with
$q_{\rm n}$ and $-q_{\rm n}$ momenta.  In fact as shown
in [7, 14, 15], there is no means to ascertain whether
the wave superposition of two particles represented
by a function like $\Psi_{\rm n}$ defines their mutual
superposition or the self superposition assumed
separately by each of them.      

\bigskip
We note that the spread ($d$) of $\Psi_{\rm n}$ is an
integer multiple of $\lambda_{\rm n}/2$ and
$\lambda_{\rm n}/2 $ equals $d$ only for the ground
state $\Psi_{\rm n=1}$.  One may call $\lambda_{\rm n}/2$
as the unit spread of $\Psi_{\rm n}$ or identify it as the
quantum size (or the uncertainty size) of a particle of
momentum $q_{\rm n}$.  It is well known that no particle
can be accommodated in a box if its $\lambda/2 > d$ which
means that no object can share the $\lambda/2$ space
occupied by a particle, particularly, if the particle and
chosen object interact through HC repulsion.  Evidently,
$\lambda/2$ represents the effective size of a quantum
particle in a sense that any effort to reduce this size
needs energy which implies that the particle {\it repels}
the object(s) trying to share the $\lambda/2$ space
occupied it.  Naturally, $\lambda/2$ could also be
identified as the range of such repulsion.

\bigskip
In the light of what follows from Section {\it 2.1} and
the fact that quantum size of a particle in its ground
state fits exactly with the size of the box, it can be
concluded that the particle has collisionless motion in
its ground state ($\Psi_{\rm n = 1}$) and collisional
motion in its excited states, $\Psi_{\rm n > 1}$.  This
agrees with similar inferences of [7, 14, 15] which imply
that collisionless motion, effective size (=$\lambda/2$)
and self superposition state are the common features of
a particle in the low energy states of widely different
many body systems where each atom could be identified
with a particle trapped in a box.

\bigskip
\centerline{\bf 3. Relevance and Experimental Support}

\bigskip
We note that the present study can help in understanding
some interesting aspects of quantum particles subjected
to certain physical situations.  For example an electron
free to move within the space of an atomic vacancy in a
1-D crystal.  We presume that: (i) the trapped electron
is unable to cross the atoms on its two sides and (ii)
the potential responsible for the forces restoring the
positions of these atoms varies as
$x^2$ (with $x =$ displacement of an atom from its
position decided by the inter-atomic interactions in the
said crystal).  While the former presumption implies that
the electron is trapped between two infinite potential
walls, the latter means that the forces restoring the
size of the trapping vacancy are identical to $F_a$.
Since this represents a case identical to the system of
present study, we use its inferences to find that the
said electron uses its $F_o$ to displace the neighboring
atoms to produce local strain in the crystal which leads
to a binding of the electron with crystalline lattice
(with binding energy $\Delta E$, Eqn. 9) and facilitates
an exchange of energy with lattice oscillations
({\it i.e.}, phonons), particularly, when $T$ of the
crystal is low enough to keep the electron in its ground
state.

\bigskip
Similarly, for two electrons occupying their ground state
in two separate atomic vacancies, one may find that their
$F_o$ should produce local strain in the crystal leading
to their binding with the lattice under the thermal
conditions which keep them in their ground state.  Both
these electrons can, obviously, be visualized to have an
{\it indirect mutual binding} (detailed mathematical
analysis given in [16]) which helps them to have a
correlated zero-point motion with an energy exchange
through phonons.  Naturally, if there are many such
electrons in the crystal, this binding can be envisaged
for each of their pair.

\bigskip
As shown in [6], we note that forces identical to $f_n$
(Eqn. 4) operate also in case of a particle trapped in a
3-D box (a cavity in a crystal).  Evidently, the
inferences of this study (Section 2.0) are relevant to
the physics of real systems like quantum dot, quantum
wire, trapped single particle/ion [8,9], {\it etc.} and
in this context, the experimentally observed $-ve$ value
of volume expansion coefficient, $(1/V)\partial_TV$ (an
analogue of $-ve$ $(1/d)\partial_Td$), for liquids $^4He$
and $^3He$ [17], respectively, around $\approx 2.2$K and
$\approx 0.6$K (not significantly different from $T_o
\approx 1.4$K estimated for these liquids), provides {\it
strong experimental support}.  This experimental fact 
also implies that the ground state of each atom in
systems like these liquids should be identical to the
ground state of a particle trapped in a 3-D box ({\it
i.e.} a cavity of neighboring particles) and $F_o$ of
each particle should, undoubtedly, have an
important role in deciding the low $T$ behavior of such
systems.  Interestingly, our recent papers, related to
microscopic theory of superconductivity [14] and
unification of the quantum behavior of widely different
systems of interacting fermions and bosons [15], clearly
prove that: (i) theories giving due importance to $F_o$
are mathematically simple, (ii) they explain several
unexplained experimental facts such as, high $T_c$ of
ceramic superconductors, superfluid $T_c$ of liquid
$^3He$, logarithmic singularity of specific heat of
liquid $^4He$ at $T_{\lambda}$, {\it etc.}, and (iii)
their results agree closely with experiments.  Evidently,
our inferences have enough experimental support and
proven relevance to the microscopic understanding of
widely different many body systems. 

\bigskip
\centerline{\bf 4. Concluding Remarks}

\smallskip
We find that : (i) a particle confined to a box assumes
a state ($\Psi_{\rm n}$) which resembles with that of
two particles in a state of equal and opposite momenta
({\it i.e.}, $q$ and $-q$), (ii) a zero-point force
($F_o$) operates when the particle occupies ground state
and produces non-zero strain ({\it viz.}, an expansion
in $d$ by $\Delta d$) since the forces restoring $d$, in
all real situations, are not infinitely strong, (iii)
the particle in its ground state has a kind of binding
with strained box, (iv) the experimental observation of
$- ve$ $(1/d)\partial_Td$ around $T = T_o$ should be an
evidence for the occurence of this strain and the fall
of particle into its ground state, and (v) the particle
motion changes from collisional to collisionless when it
moves from its excited state ($\Psi_{\rm n > 1}$) to
ground state ($\Psi_1$).  Since these results are as
natural consequences of wave particle duality as Eqns.
1-4, the experimentally observed $-ve$
$(1/V)\partial_TV$ for liquids $^4He$ and $^3He$
(respectively, around 2.2 and 0.6 K) unquestionably
support the above mentioned inferences and this implies
that: (i) mechanical strain resulting from $F_o$, (ii)
$-ve$ expansion coefficient around $T_o$ and (iii) loss
of relative motion leading to collisionless motion of
quantum particles ({\it as concluded in [7] with more
detailed discussion}) should be common aspects of the
low energy states ({\it close to ground state}) of
widely different many body systems. It may, therefore,
be emphasized that a microscopic theory of a many body
system, like electron fluid, liquids $^4He$ and $^3He$,
{\it etc.}, can not truly explain their low $T$
behavior unless it incorporates the role of $F_o$
({\it a basic consequence of the wave particle duality})
and related aspects.  As a proof of this point one may
find recent studies related to the basic foundations of
microscopic theory of superconductivity [14] and
unification of the physics of widely different systems
of interacting bosons and fermions such as liquids
$^4He$ and $^3He$ [15].

\bigskip
This study should be useful for the learners of wave-
mechanics because it reveals some of the basic aspects
of a particle in a box ({\it e.g.}, the consequences of
$F_o$) for the first time and to this effect it
deserves to be a part of the elementary texts on wave
mechanics. It should also be interesting to those who
desire to contribute towards a true and complete
understanding of the low energy states of widely
different many body systems and the exotic phenomena of
superfluidity and superconductivity.

\bigskip
\noindent
{\it Acknowledgement} : The author is thankful to
Dr. R. Moro (School of Physics, Georgia Institute of
Technology, Atlanta, USA) for his very useful
suggestions.

\end{document}